\documentclass{bmcart}
\usepackage[utf8]{inputenc} 
\usepackage{graphicx}
\usepackage{hyperref}
\usepackage{booktabs}
\usepackage[colorinlistoftodos]{todonotes}
\usepackage[all]{nowidow}
\usepackage{url}

\startlocaldefs
\endlocaldefs

\newcommand{\overbar}[1]{\mkern 1.5mu\overline{\mkern-1.5mu#1\mkern-1.5mu}\mkern 1.5mu}

\begin{document}

\begin{frontmatter}

\begin{fmbox}
\dochead{Research}


\title{Shopping Mall Attraction and Social Mixing at a City Scale}



\author[
   addressref={aff4},
   email={mbeiro@fi.uba.ar}
]{\inits{MB}\fnm{Mariano G.} \snm{Beir\'o}}
\author[
   addressref={aff1,aff2},
   email={bravo@udd.cl}
]{\inits{LB}\fnm{Loreto} \snm{Bravo}}
\author[
   addressref={aff1,aff2},
   email={dcaro@udd.cl}
]{\inits{DC}\fnm{Diego} \snm{Caro}}
\author[
   addressref={aff3},
   email={ciro@isi.it}
]{\inits{CC}\fnm{Ciro} \snm{Cattuto}}
\author[
   addressref={aff1,aff2},
   email={lferres@udd.cl},
]{\inits{LF}\fnm{Leo} \snm{Ferres}}
\author[
   addressref={aff1,aff2},                   
   email={egraells@udd.cl}   
]{\inits{EG}\fnm{Eduardo} \snm{Graells-Garrido}}

\address[id=aff4]{%
  \orgname{Facultad de Ingeniería, Universidad de Buenos Aires, INTECIN (CONICET)},
  \street{Av. Paseo Colón 850},
  \postcode{C1063ACV}
  \city{Buenos Aires},
  \cny{Argentina}
}
\address[id=aff1]{
  \orgname{Data Science Institute, Faculty of Engineering, Universidad del Desarrollo}, 
  \street{Av. La Plaza 680, Las Condes},                     %
  \postcode{}                                
  \city{Santiago},                              
  \cny{Chile}                                    
}
\address[id=aff2]{%
  \orgname{Telef\'onica R\&D},
  \street{Av. Manuel Montt 1404, Third Floor, Providencia},
  \postcode{}
  \city{Santiago},
  \cny{Chile}
}
\address[id=aff3]{%
  \orgname{ISI Foundation},
  \street{ Via Chisola 5},
  \postcode{10126}
  \city{Torino},
  \cny{Italy}
}

\end{fmbox}


\begin{abstractbox}

\begin{abstract} 
The social inclusion aspects of shopping malls and their effects on our understanding of urban spaces have been a controversial argument largely discussed in the literature. Shopping malls offer an open, safe and democratic version of the public space. 
Many of their detractors suggest that malls target their customers in subtle ways, promoting social exclusion. In this work, we analyze whether malls offer opportunities for social mixing by analyzing the patterns of shopping mall visits in a large Latin-American city: Santiago de Chile.

We use a large XDR (Data Detail Records) dataset from a telecommunication company to analyze the mobility of $387,152$ cell phones around $16$ large malls in Santiago de Chile during one month. We model the influx of people to malls in terms of a gravity model of mobility, and we are able to predict the customer profile distribution of each mall, explaining it in terms of mall location, the population distribution, and mall size.

Then, we analyze the concept of social attraction, expressed as people from low and middle classes being attracted by malls that target high-income customers. We include a social attraction factor in our model and find that it is negligible in the process of choosing a mall. We observe that social mixing arises only in peripheral malls located farthest from the city center, which both low and middle class people visit. Using a co-visitation model we show that people tend to choose a restricted profile of malls according to their socio-economic status and their distance from the mall. We conclude that the potential for social mixing in malls could be capitalized by designing public policies regarding transportation and mobility. 
\end{abstract}





\end{abstractbox}
%

\end{frontmatter}



\section{Introduction}





Shopping malls have a prominent place in the configuration of modern cities, affecting the daily activities, social relationships and mobility of their inhabitants. They arose in the U.S. during the postwar, and have been idealized since then as democratic spaces to which all citizens have equality of access.
However, this ideal has proven to be false in the U.S., in which malls are essentially a space for white middle classes~\cite{goss1993magic, staeheli2006usa, lofland2017public}. By being located in suburban
areas or areas without adequate public transportation, they become unreachable for the lower-middle class, for example people without a car, thus promoting minority exclusion and segregation~\cite{austin1997not}.

Though many shopping malls are currently dying in the U.S.~\cite{decline}, their concept has been replicated throughout the world, and in many cities they are strong constituents of the urban space, specially in developing countries.

The aim of our work is to assess the potential of shopping malls for social mixing and inclusion. We will focus our work in Chile, were we used Data Detail Records (XDRs) provided by Telef\'onica R\&D to analyze the mobility patterns of people going to malls in its capital city, Santiago de Chile, in order to determine which factors influence mall choice and what types of social mixing can be found in malls. As far as we know, malls in Latin America have never been studied using large populations by means of CDR (Call Detail Records) or XDR data.

Shopping malls became popular in Chile in the early 80's, and were one of the first signs of the globalization and liberalization of consumption in Latin America. They are currently one of the main elements of Chilean popular culture, influencing people's mobility, purchase decisions, self-expression, and segregation processes~\cite{Salcedo2013}. 
Furthermore, in comparison with other emerging countries, Chile is the Latin-American country with the highest mall surface rate per inhabitant~\cite{eyn}.

In Santiago de Chile, as in many other large cities from developing countries, socio-economic class segregation has a different aspect than in the U.S.: on the one hand, the extensive public transportation system allows malls' close integration with pedestrian and city life; on the other hand, many low income people live in peripheral areas, and the arrival of malls to these zones has permitted the integration of middle- and low-income consumers~\cite{stillerman2012transposing}. These situations allow for cross-class encounters in malls. We put this idea to test by employing a recent segregation framework~\cite{louf2016patterns}, and find that, indeed, there is social mixing -- but not for everyone. Then we ask whether social mixing plays a role in mall selection,\textit{i.e.}, if people might choose a mall partly motivated by the opportunity of mixing with people of other classes. In order to understand the factors behind this selection, we fitted a gravity model of travel which explains the aggregated mobility flows of people from different city areas to malls in terms of population, mall size, and distance. We observed that this model can almost perfectly predict the socio-economic distribution of the customers visiting each mall. We verified that the attractiveness of malls in our model obeys O'Reilly's law of retail gravitation~\cite{reilly1931law}.

After fitting the gravity model, we turned into a more individual model of mall selection in which we fitted the probability of a customer visiting some mall $B$ given that he visited mall $A$. From this model we observed that individual selection is not only conditioned by distance and mall size, but also by the similarity between malls: \textit{i.e.}, people visiting a high target mall will most probably visit another high target mall rather that a low-middle class mall. However, we observed that low and middle-class people do mix in shopping malls around the periphery of the city, but few of these people reach distant, high-target malls.
Our results suggest that social mixing and inclusion can be promoted by improving accessibility and public transportation in order to reduce travel times for low and middle classes.

\section{Related work}

The effects of malls in urban societies in the U.S. have been extensively studied in the literature of the social sciences. Among the most prominent works, L. Cohen studied their emergence in the context of suburbanization~\cite{cohen1996town}; J. Goss depicted them as new civic spaces that manipulate behavior through the configuration of space, in order to provoke dispositions and facilitate consumption~\cite{goss1993magic}; Wakefield and Baker have applied factor analysis to identify the determinants of customer excitement~\cite{wakefield1998excitement}.
O'Reilly measured the effects of mall size and distance on mall selection and proposed the Law of Retail Gravitation: he suggested that larger retail centers present a higher attraction to customers, who would be willing to travel longer distances to arrive to them~\cite{reilly1931law}; Huff studied the concept of trading area and proposed a model for defining the catchment area of a retail center~\cite{10.2307/1249154}.

Currently, large shopping malls in the U.S. are suffering a decline since $2005$, partly due to the increase of online retail, but also to the proliferation of convenience and neighborhood retail as part of the new Smart Growth movement that is reshaping many US cities~\cite{decline}. Supporters of Smart Growth promote a fine-grained scale integration of housing, jobs and retail spaces, suggesting that mixing land uses and prioritizing open spaces in neighborhoods will promote inclusion and social mixing~\cite{smart_growth}. 

In a Latin-American context, A. Dávila has analyzed the effects of the access to shopping malls for the new middle class~\cite{10.1525/j.ctt19632h7}. In a case study in Bogotá (Colombia) she observed that, instead of reducing social differentiation, malls reproduce the cultural hegemony of a living traditional elite. Stillerman and Salcedo studied how people use and understand malls in Santiago de Chile, observing cross-class encounters in two malls in the periphery of the city~\cite{stillerman2012transposing}. He concluded that people use malls to satisfy different types of needs: from purchasing to self-expressing and keeping familiar, romantic and social relationships. Dinzey-Flores analyzed the antagonism between malls and poverty in Puerto Rico~\cite{dinzey2017spatially}, which has also been seen (more violently) in Brazil~\cite{goncalves2014conflicting}.

Retail location has also been studied in the context of urban mobility: ~\cite{rob} have studied the impact of workplace proximity, housing, and retail accessibility on mobility, showing that job-housing proximity has a larger impact on total mobility than retail location. 
On the other side, Galati and Greenhalgh studied the inner mobility, contact duration and inter-contact time of mall customers~\cite{Galati:2010:HMS:1755743.1755745}. In the context of big-data marketing, Chen~{\em et al.} have used a large dataset for optimizing retail store location~\cite{chen2018geographic}. 

\section{\label{secMethods}Methods}

In this section we will introduce the three methods we will use in this article to measure social mixing, study the attraction to malls and cluster them according to their costumer profiles.  

\subsection{Measuring Social Mixing}

Several models of segregation exist, yet, recently Louf and Barthelemy~\cite{louf2016patterns} proposed a new one that consideres a null model of spatial segregation. In it, the exposure of a population $\alpha$ to a population $\beta$ is defined as:

\begin{equation}
E_{\alpha\beta} = \frac{1}{N_\alpha} \sum_{m = 1}^{M} n_{\alpha}(m) \times r_{\beta}(m)
\label{eq:e_a_b}
\end{equation}

\begin{equation}
r_{\beta} (m) = \frac{n_\beta(m) / N_\beta}{n(m) / N}
\label{eq:r_a}
\end{equation}

Where, in our context:

\begin{itemize}
    \item $N_\alpha$ is the total number of people in category $\alpha$.
    \item $\{1,\dots,M\}$ is the set of malls.
    \item $n_{\alpha}(m)$ is the total number of visitors to mall $m$ belonging to category $\alpha$.
    \item $r_\alpha(m)$ is the \emph{representation} of category $\alpha$ in mall $m$ as computed by formula (\ref{eq:r_a}).
    \item $n(m)$ is the total number of visitors to mall $m$.
    \item $N$ is the total number of people. 
\end{itemize}

\noindent In summary, ``the representation compares the relative population $\alpha$ in the areal unit $m$ to the value that is expected in an unsegregated city where individuals choose their location at random.'' \cite{louf2016patterns}. To apply this definition, we segment users into categories by analyzing their socio-economic characteristics, and binning them into quantiles. The exposure metric is interpreted as follows: if $E_{\alpha\beta} >$ 1, then social mixing happens  between subpopulations $\alpha$ and $\beta$ within malls. Conversely, if  $E_{\alpha\beta} <$ 1, then both categories are segregated. 

Our hypothesis is that malls function as social aggregators, and thus, they promote social mixing in terms of colocations people from different socioeconomic status in the same spaces. Even though colocation is not real interaction, it provides a venue for \emph{potential interactions} that, in terms of the model, are statistically significant in comparison to the null model of random encounters \cite{louf2016patterns}.

\subsection{A gravity mobility model for mall visits}

If malls are social enablers, then the next question is whether this phenomena influences why people choose to go to specific malls. To measure this behavior, we resort to the gravity model of mobility.

Different human mobility models have been proposed in the literature to explain the collective flows of people in the context of commuting, migration or tourism:  being the gravity model and the radiation model the most used.

The gravity model of flow was proposed by W. Alonso~\cite{alonso1976theory} for modelling spatial flow phenomena between regions. It has been extensively used in the literature for modelling international air travel~\cite{grosche2007gravity}, traffic flow in highways~\cite{highways_2008} and telecommunications across cities~\cite{krings_calls}, among others.

The more recent radiation model~\cite{simini2012universal} has proven to outperform the gravity model for commuting flows across cities. As an advantage it does not require real mobility data for fitting any parameters, as the flows are predicted entirely from population distributional data. However,  its performance has been inferior for intra-urban mobility~\cite{liang2013unraveling, palchykov2014inferring}.

In this work we shall model the movements of people living from different areas of a city to a set of large malls using the gravity model. Our selection is based on several reasons: first, the gravity model has a good performance at the intra-urban level; secondly, the same gravity law has been used for estimating the flow of economic goods between countries~\cite{tinbergen1962shaping} and the attractiveness of retail centers. In particular, O'Reilly's law of retail gravitation models the attractiveness as directly proportional to the size of the retail center and inversely proportional to the square root of its distance~\cite{reilly1931law}. This law has been generalized later in terms of the gravity model~\cite{batty1978reilly}. However, O'Reilly's model does not take cultural differences into account: it might be the case that people from certain social groups prefer certain types of malls instead of others. One of our hypothesis is that cultural differences are strong determinants of the social mixing that emerges in malls. If that were the case, then we should see that social mixing cannot be exclusively explained by distances, mall sizes, and O'Reilly's law. We will assess the performance of a generalized gravity model for predicting influx of people to malls, with and without considering a social attraction factor.

In an analogy with Newton's gravity law, the gravity model of flow considers the flow between two nodes $(i, j)$ as directly proportional to the sum powers of both nodes' populations, and inversely proportional to a power of the distance between them:

\begin{equation}
F_{ij}=G \frac{M_i^{\alpha}M_j^{\beta}}{D_{ij}^{\gamma}},
\label{gravityF}
\end{equation}


\noindent where:

\begin{itemize}
\item $M_i$ is the population of a square grid in the city, computed from census data;
\item $M_j$ is the mall size in terms of total rental space;
\item $D_{ij}$ is the distance in kilometers between the center of the square grid and the mall;
\end{itemize}


\noindent The traditional approach for fitting this model consists in applying a logarithmic transformation to the equation, leading to a linear model on the logarithms of the variables:

\begin{equation}
log(F_{ij}) = log(G) + \alpha\cdot log(M_i) + \beta\cdot log(M_j) - \gamma\cdot log(D_{ij}) + \epsilon_{ij}
\label{gravityF}
\end{equation}

\noindent where $\epsilon_{ij}$ represents an aditive, independent error term. This linearlized model can be fitted through OLS (ordinary least squares)~\cite{goldberger1968interpretation}. However, this approach has several limitations: it cannot model the zero observations (which must be thrown away), and the estimated coefficients can have significant biases under heteroskedasticity~\cite{silva2006log}. As an alternative, we replace the linear regression by a Poisson one (a particular case of the Generalized Linear Model, (GLM)~\cite{mccullagh1984generalized}):

\begin{equation}
\mathbf{E}[F_{ij}] = exp[\log(G) +\alpha\cdot log(M_i) + \beta\cdot log(M_j) - \gamma\cdot log(D_{ij})]
\label{espgm1}
\end{equation}

\noindent This new version is fitted by maximizing the log-likelihood function. The maximization of this function does not have a closed analytical solution, but as the function is convex convergence is guaranteed by applying standard optimization techniques such as gradient descent or iteratively reweighted least squares (IWLS).

Additionally, we shall consider an augmented gravity model in which we consider an attraction factor $A_{ij}$ representing the difference between the average customer profile of mall $j$ and the Human Development Index of square $i$ ($\mathit{HDI}_i$):

\begin{equation}
\mathbf{E}[F_{ij}] = exp[\log(G) +\alpha\cdot log(M_i) + \beta\cdot log(M_j) - \gamma\cdot log(D_{ij}) - \lambda\cdot log(A_{ij})]~~~~~
\label{espgm2}
\end{equation}

\noindent where $A_{ij} = \overbar{\mathit{HDI}}_j - \mathit{HDI}_i$. In this last equation, a mall's average customer profile $\overbar{\mathit{HDI}}_j$ is defined as the mean $\mathit{HDI}$ value of all the people visiting mall $j$ during the month.

\subsection{Clustering malls according to customer profiles}


In order to better understand the motivations behind mall selection, we built a co-visitation network representing common mall customers. This is a weighted directed network whose nodes $v_i$ are the $16$ malls, while the weighted edges $(v_i, v_j)$ between them represent the conditional probability of visiting mall $v_j$ given that someone visited mall $v_i$. We built a similarity matrix $S$ between malls, by computing the Kolmogorov-Smirnov $S_{ij}$ distance between the customer profile distributions of the malls. 

Then, we built a Logit model for predicting the conditional probability of visiting mall $v_j$ given that a customer visited mall $v_i$. We fitted this model using a logistic regression under the following specification:

\begin{equation}
\mathbf{E}[p_{j|i}] = (1 + exp[-log(K) - \beta\cdot log(M_j) - \lambda\cdot log(S_{ij}) + \gamma\cdot log(D_{ij})])^{-1}
\label{espcm}
\end{equation}




\section{Datasets and Data Preprocessing}
\label{sec:datasets}

\begin{figure}[t]
\centering
    \includegraphics[width=0.65\linewidth]{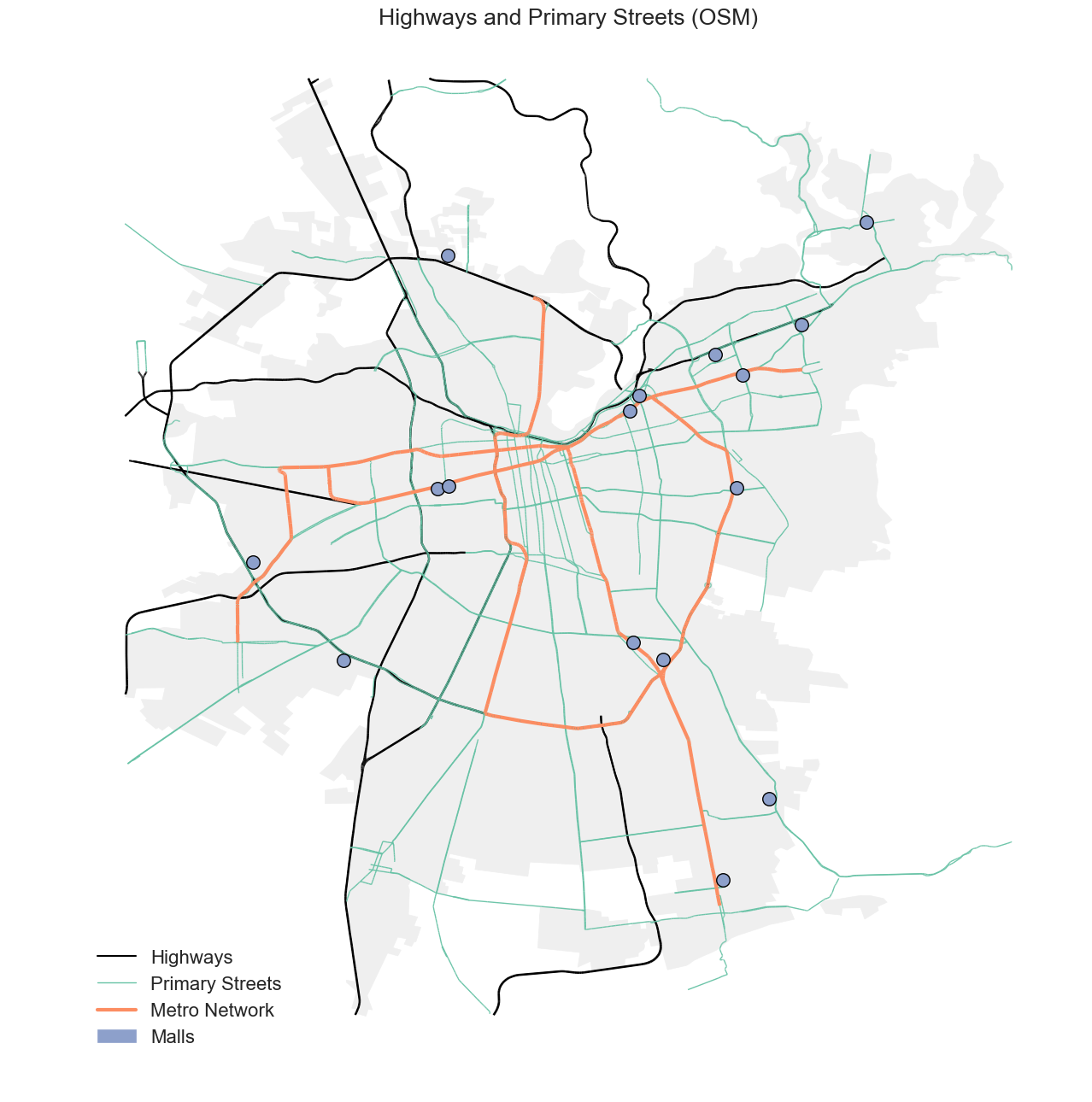}
    \caption{\label{fig:santiago}Schematic map of Santiago, urban area. Lines encode highways (black), primary streets (green), the Metro network (orange). Markers are the malls under study. The urban network data is from OpenStreetMap.}
\end{figure}

\begin{table}[t]
      \begin{tabular}{clrr}
\hline
Id & Mall & Rental Space (sqm) & Visitors \\
\hline
\hline
1	& Alto las Condes	& 115,000 & 122,061\\
2	& Apumanque	& 20,100 & 36,154\\
3	& Mall Arauco Maipú	& 54,000 & 94,788\\
4	& Mall Costanera Center	& 173,000 & 163,234\\
5	& Espacio Urbano Puente Alto & 25,500 & 77,153\\
6	& Mall Florida Center	& 110,000 & 145,436\\
7	& Mall Vivo Panorámico	& 7,000 & 69,646\\
8	& Mall Parque Arauco	& 111,000 & 170,494\\
9	& Mall Paseo Estación	& 46,700 & 59,056\\
10	& Mall Plaza Alameda	& 60,000 & 26,478\\
11	& Mall Plaza Egaña	& 80,000 & 58,239\\
12	& Mall Plaza Norte	& 74,900 & 127,768\\
13	& Mall Plaza Oeste	& 144,000 & 148,716\\
14	& Mall Plaza Tobalaba	& 54,200 & 2,070\\
15	& Mall Plaza Vespucio	& 116,000 & 96,523\\
16	& Portal La Dehesa	& 79,000 & 29,268\\
\hline
      \end{tabular}
\vspace{0.2cm}
\caption{\label{tbl:mallsvisits}Malls under study in Santiago, with their size in square meters (according to Chile's Chamber of Commerce), and number of unique visitors in our dataset during August, 2016.}
\end{table}

Our study focuses in $16$ malls located in the urban area of Santiago, Chile, taken from \cite{Salcedo2013}. Table \ref{tbl:mallsvisits} shows the rental space of these malls, and Figure \ref{fig:santiago} shows them in the urban context, in terms of the urban network of streets and transportation services. The urban area is comprised by 35 \textit{``comunas''}.\footnote{We have decided to maintain the Spanish word for the administrative unit of Chile: ``county'' or ``commune'' being close translations to English, but rather artificial.} 
In total, the urban area of the city comprises 35 \textit{comunas}. 


The Santiago urban area contains $13,239$ cellphone antennas, distributed in $1,377$ towers. Some antennas are located indoors, such as in underground metro stations, important public and private buildings, including, interestingly, malls, which reflects the importance of this kind of building. As these antennas are usually low power, connections to them are almost surely established from inside the mall. It is very improbable that a passerby can connect to an inner antenna from outside, as it is also improbable that a mall visitor will establish a connection to some antenna outdoors. Antennas inside malls could be easily identified in two different ways: {\em (i)} by means of a description field in the dataset which is human readable and contains the word ``mall'' in it; {\em (ii)} by drawing each mall's polygon and using the towers' coordinates to determine which ones were inside the polygons.\footnote{The mall polygons are available at \url{https://github.com/leoferres/mallmob}} Both methods turned out to be highly consistent. This yielded 481 indoor cellphone antennas placed \textit{inside} the malls of interest, which is $29.25\%$ of all indoor antennas in the Santiago urban area, highlighting the important role of shopping malls in Santiago.

The dataset consists of $1,023,118$ unique anonymized mobile devices (containing a Movistar/Telef\'onica SIM card) triggering events with the indoor mall antennas and interacting with the Telef\'onica network in the course of one single month: August 2016. Each XD Record contains information logged by Telef\'onica in Santiago during that month, including latitude/longitude pairs, hashed phone number, time-stamp, and bytes downloaded \cite{Graells-Garrido2017}. The XDR (rather than the more common CDR or Call Detail Record) stores interactions for, for example, accesses to websites, applications that communicate over the Internet \cite{Calabrese:2014:USU:2658850.2655691}. As such, XDRs only have one communicating antenna, (CDRs have two instead: one for the caller and one for the callee), and the cost is measured in down/upload units (bytes) since the last registered event rather than minutes consumed. Given the prevalence of data rather than voice-driven communication nowadays (interacting with the internet is more common than calling someone), XDRs allow much more time granularity, at around 15--30 minutes between logged events. Calls are more sparse.

\begin{figure}[t]
\centering
    \includegraphics[width=0.75\linewidth]{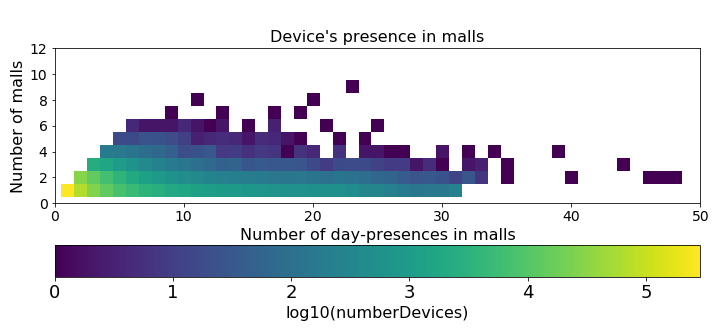}
    \caption{\label{fig_dev_presence}Histogram representing the number of devices that performed a certain number a mall visits during the month ($x$-axis), and whose visits spanned a certain number of different malls ($y$-axis).
    }
\end{figure}

Figure~\ref{fig_dev_presence} shows a histogram of the devices that were found as visiting a certain number of different malls during the month ($y$-axis), and totalizing a certain number of visits ($x$-axis), a visit being defined as the presence of the device inside a mall during a specific day. For the $81,027$ devices with more than 10 day-presences in malls, we conservatively determined that they were not mall customers and discarded them from the dataset (the are probably related to SIM cards associated to card readers, and to people who are not visitors, such as employees or providers). We identified $942,091$ devices (of the original $1,023,118$) as ``real visitors.'' The remaining ones might be associated with staff, mall providers or utility devices such as credit card readers, points-of-purchase, \textit{etc.} The total number of visitors per mall is shown on Table \ref{tbl:mallsvisits}.

\begin{figure}[t]
\centering
    \includegraphics[width=0.85\linewidth]{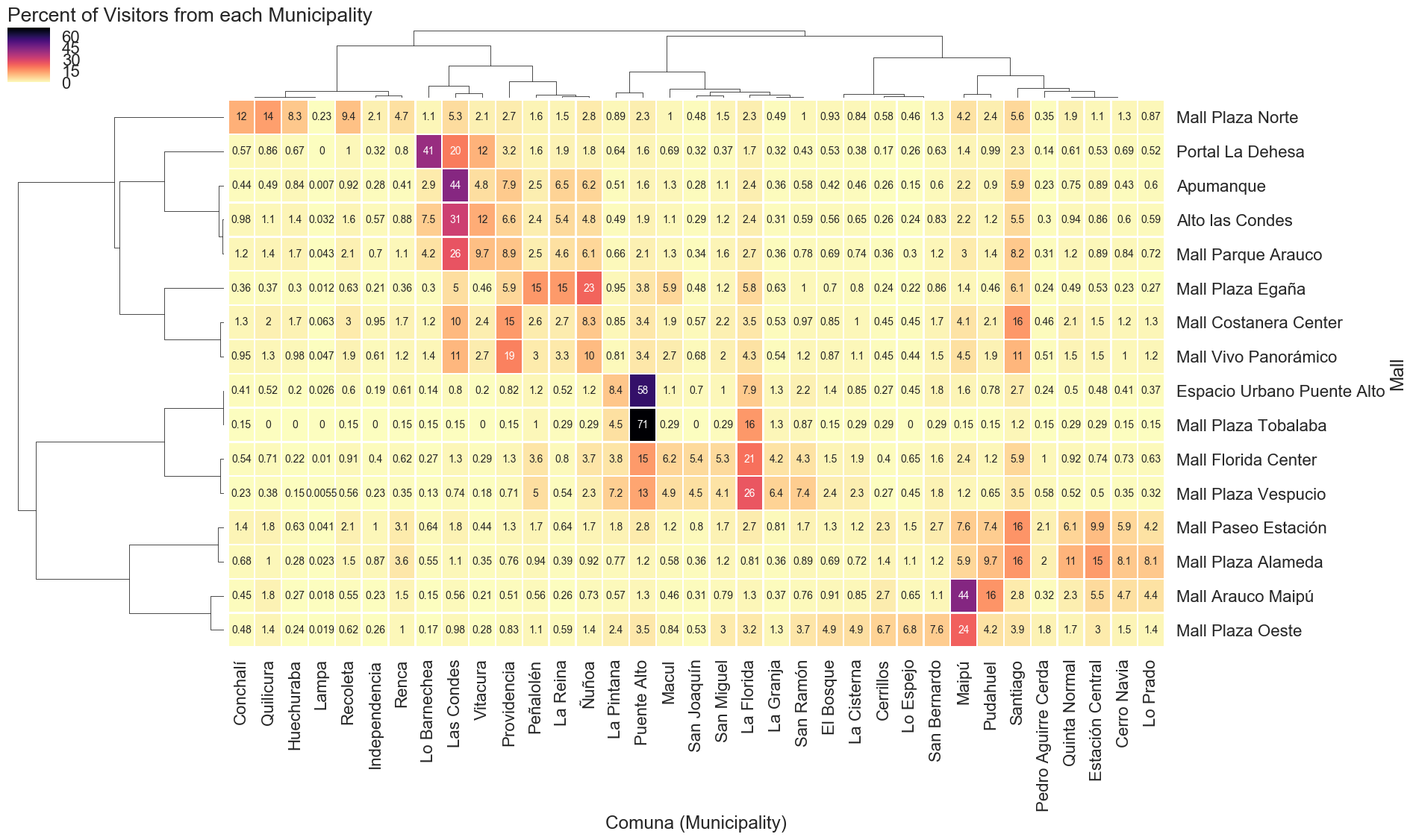}
    \caption{\label{fig:municipality_mall_matrix}Heatmap of mall visitors according to their \textit{comuna} of residence. The matrix is row-normalized, meaning that each row encodes the percent of mall visitors from each \textit{comuna}. Note that, even though the city is named Santiago, the center \textit{comuna} is also named Santiago.}
\end{figure}

For those devices identified as visitors we determined the area where they live (the ``home antenna'') by finding the first (before 8AM) and last (after 10PM) cell tower they were connected to each day of the month. We then computed, for each user: 
{\em (i)} The number of times in which we detected a first/last tower connection, 
and {\em (ii)} the relative frequency of the most observed tower (\textit{i.e.}, the number of times in which the user's first or last tower was that tower, divided by the previous quantity). Results show that more than 70\% of the devices were connected at least 80\% of days. About half the users repeated their most common first-last tower at least 60\% of the days in which they are seen. In order to ensure a good confidence for the home location determination, we only kept these last consumers ($387,152$ unique devices). A Pearson correlation coefficient was computed to asses the relationship between our home-antenna identification approach and the actual population of those \textit{comunas}, given by the 2017 census information \cite{censo17}, obtaining a positive correlation of $0.84$. Although a simple method for calculating home-antenna (\textit{c.f.} more sophisticated ones like \cite{ALEXANDER2015240,graells2016sensing}), the results are strong enough for our purposes.

The heatmap in Figure \ref{fig:municipality_mall_matrix} shows the distribution of the \textit{comunas} of residence of mall visitors. One can see that several malls are tailored at specific \textit{comunas} due to the high fraction of visitors from one or two (\textit{e.g.}, Mall Arauco Maipú exhibits a high fraction of visitors from Maipú; Mall Plaza Tobalaba exhibits a high fraction of visitors from Puente Alto), while others exhibit high diversity (\textit{e.g.}, Mall Plaza Norte, Costanera Center, Panoramico, Mall Plaza Vespucio, \textit{etc.}). This implies that, in terms of the \textit{comuna} of residente, there may be social mixing in some malls.

\begin{figure}[t]
\centering
    \includegraphics[width=0.65\linewidth]{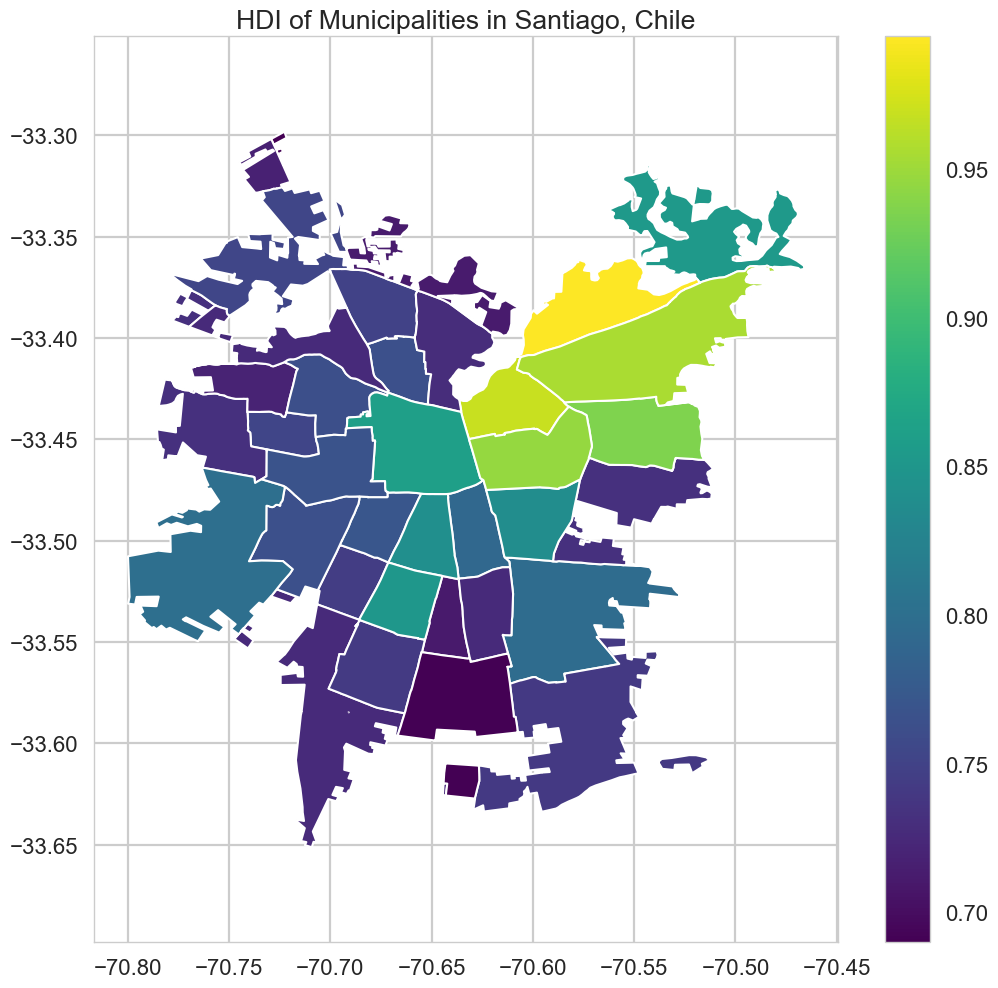}
    \caption{\label{fig:santiago_hdi}Choropleth map of Santiago. The color scale represents the estimated Human Development Index (HDI) of each \textit{comuna} within the city.}
\end{figure}

\begin{figure}[t]
\centering
    \includegraphics[width=0.95\linewidth]{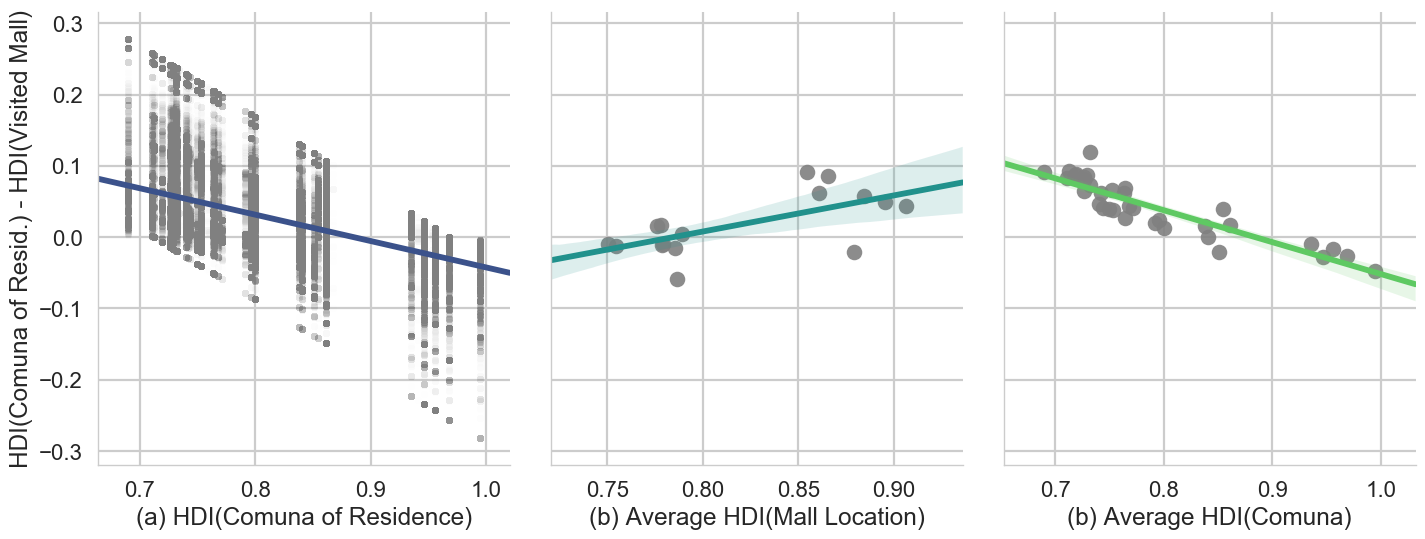}
    \caption{\label{fig:municipality_hdi_differences}Correlations of HDI: (a) \textit{comuna} of residence, and the difference between HDI of the \textit{comunas} of visited malls; (b) HDI Average HDI of each \textit{comuna}'s inhabitants, and the difference with the HDI of their visited malls; (c) Mall HDI (based on mall location), and the average differences between visited mall HDI and HDI of user's residence.}
\end{figure}

A further dataset needed for the purposes of this work was a measure of the Human Development Index, or HDI. Unfortunately, in Chile, the last formal/official calculation of the HDI at the \textit{comuna} level was done in 2005~\cite{Programa10798}
. To make it more up to date we computed the HDI using the new method proposed by the United Nations in 2010~\cite{klugman2010human}, using open governmental data on income distribution, life expectancy and education for Chile for the period 2013-2015.\footnote{Data from CASEN: \url{https://goo.gl/LuE6s3}, source code at~\url{https://goo.gl/FjvWpW}} We correlated our results against those of 2005, obtaining a Pearson correlation of $0.90$, making them comparable. We use our HDI calculation in the rest of this paper. Figure \ref{fig:santiago_hdi} shows the geographical HDI distribution, putting into evidence the socio-economic segregation of the city.

One direct way to explore social mixing is to aggregate mall visits per user, estimate the difference between HDI of visited mall (in terms of a mall's location) and HDI of \textit{comuna} of residence, and then calculate the correlation of those differences with the HDI of each \textit{comuna}. Figure \ref{fig:municipality_hdi_differences} shows three correlations of such differences: (a) individual differences (Pearson's $r =$~-0.49, p~$< 0.001$; (b) aggregated per \textit{comuna} of residence ($r =$~-0.9, p~$< 0.001$); and (c) aggregated per visited mall ($r =$~0.65, p~$< 0.01$). In all cases, the difference was normalized by the number of days each user visited a mall. These results support our motivation: on the one hand, users have a tendency to visit malls in areas with higher HDI than their residence's; on the other hand, malls tend to receive visitors from areas with lower HDI. However, the \textit{comuna} level is too coarse to bring conclusions in this matter. In the next section we evaluate exposure to people with different HDI using the theory described in the methods section, at a finer granularity.

\section{Results}

\begin{table}[t]
\centering
\begin{tabular}{llr}
\toprule
Quantile & HDI Range &  \# of Users \\
\midrule
Q1 & (0.689, 0.732] &     56,863 \\
Q2 & (0.732, 0.764] &     50,335 \\
Q3 & (0.764, 0.8]   &     52,921 \\
Q4 & (0.8, 0.946]   &     57,585 \\
Q5 & (0.946, 0.995] &     48,164 \\
\bottomrule
\end{tabular}
\caption{Results of quantization of HDI values of users.}
\label{tbl:hdi_quantiles}
\end{table}

\begin{figure}[t]
\centering
    \includegraphics[width=0.95\linewidth]{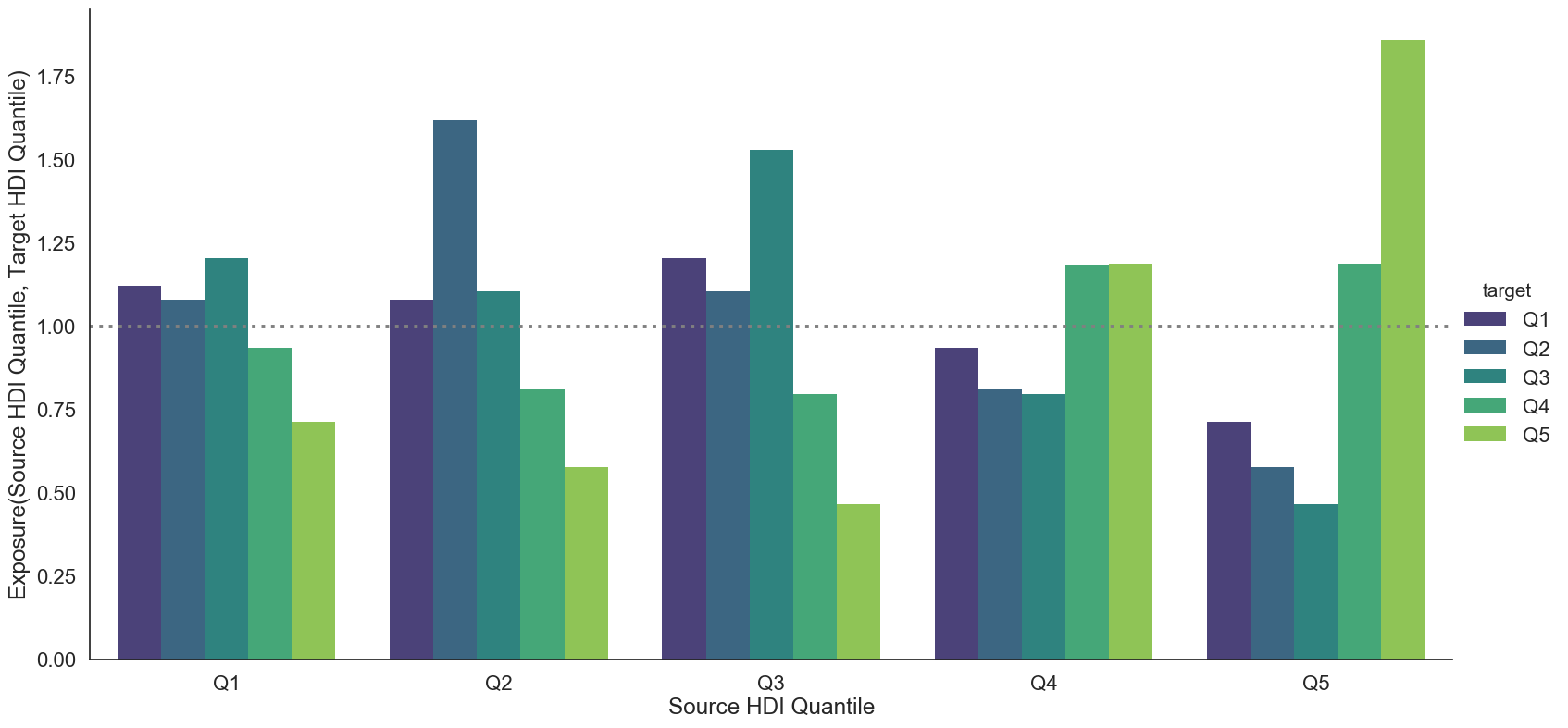}
    \caption{\label{fig:hdi_exposure}Results of the social mixing estimation, using the segregation model \cite{louf2016patterns}. Each bar shows the value of social mixing $E$ for two categories of HDI. A value of $E >$~1 implies social mixing, while a value of $E <$~1 implies social exclusion. All $E$ values are significant in this figure.}
\end{figure}

\subsection{Social Mixing seen through Segregation Theory}


Since the model by \cite{louf2016patterns} requires the definition of categories, we binned users by estimating the quantiles of the HDI distribution, assigning to each user the HDI of their \textit{comuna}. Table \ref{tbl:hdi_quantiles} shows the results.

Then, we evaluated $E_{\alpha\beta}$, where $\alpha$ and $\beta$ where the HDI quantiles. In Figure \ref{fig:hdi_exposure}, which displays the several values of $E_{\alpha\beta}$ for all possible HDI pairs, one can see that there is varying behavior in the value of $E$ (all $E$ values are significant with 99\% confidence according to the variance estimation in \cite{louf2016patterns}).

We observe that the first three quantiles (Q1, Q2, and Q3) tend to have a value of $E >$ 1 between them, and $E <$ 1 with the other two quantiles (Q4 and Q5). However, the opposite trend happens for quantiles Q4 and Q5. This implies that social mixing happens, but with restrictions -- people co-locate with others who are in ``nearby'' quantiles of HDI.

When $\alpha = \beta$, the $E$ metric is called the \emph{isolation} of $\alpha$: $I_alpha$. A value $I_\alpha >$ 1 implies that the $\alpha$ groups tends to co-locate with people from the same group. Three groups show high values of isolation: Q2, Q3 and Q5. Q5 is the most interesting case, as it presents the highest isolation of all, low values (\textit{e.g.}, exclusion) with the lower three quantiles. 

These results imply that there is social mixing in malls in Santiago, Chile. Such co-location happens at several (but not all) levels, suggesting that there is segregation in mall visitors, which may be analogous to the known socio-economical segregation of Santiago, Chile. 

\subsection{Fitting the Gravity model}

Next, we fitted a gravity model of flow between areas of the city and malls. Since there is social mixing, and our hypothesis states that people looks for this co-location, we hypothesized that the difference in HDI between a user's residence and the malls they visited should have a significant weight on the model.

However, even though HDI is estimated at \textit{comuna} level, to fit and evaluate the model we needed a finer spatial granularity. To do so, we aggregated \textit{origin} points are aggregations of antennas at a radius of 1.1KM (obtained by clipping the latitude and longitude of the real antennas to two decimals), effectively creating a grid with an ``abstract'' antenna in the middle of each square in the grid. In turn, our destination points are the center of the polygon malls mentioned in Section \ref{sec:datasets}. The population of each square in the grid is calculated using the whole 2012 Census, with information at the level of blocks. In this case, we calculate population at the level of census zones which aggregate the information of a few blocks. We also calculated the ``mass'' of a given mall using the total rental space of the mall, with data taken from the Chamber of Commerce of Chile \cite{ccch}. Using the XDR dataset, the model was calibrated by extracting the number people living at each abstract antenna $C$ and visiting each mall $M$. As discussed above, a person is said to visit mall $M$ if that person has been there at least once during the day, with the provisos made above to rule out non-visitor events. Likewise, we assume that a visitor lives in an abstract antenna $C$ when that person's home antenna is within the boundaries of the square represented by $C$.

We summarize the results of the Poisson regression for our gravity model in Table~\ref{tab_poisson}. We find exponent values of $\alpha=0.52$, $\beta=0.49$ and $\gamma=1.16$. In particular, from the fact that $\gamma/\beta=2.34$ we observe that our results obbey O'Reilly's law of retail gravitation, which suggests that $1.5 \leq \gamma/\beta \leq 2.5$.

\begin{figure}[h!]
\centering
    \includegraphics[height=7cm]{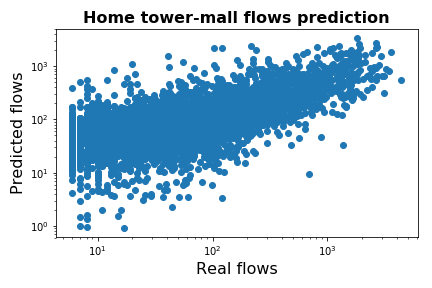}
    \caption{\label{scatterp}Scatter plot representing predicted flows vs. real flows from each cell of the square grid to each mall.}
\end{figure}

\begin{table}[]
    \centering
    \begin{verbatim}
                     Generalized Linear Model Regression Results                  
==============================================================================
Dep. Variable:                   F_ij   No. Observations:                 5488
Model:                            GLM   Df Residuals:                     5484
Model Family:                 Poisson   Df Model:                            3
Link Function:                    log   Scale:                             1.0
Method:                          IRLS   Log-Likelihood:            -1.8089e+05
                                        Deviance:                   3.3302e+05
                                        Pearson chi2:                 4.87e+05
==============================================================================
                 coef    std err          z      P>|z|      [0.025      0.975]
------------------------------------------------------------------------------
logG          -3.6642      0.036   -101.966      0.000      -3.735      -3.594
alpha          0.5240      0.003    209.410      0.000       0.519       0.529
beta           0.4944      0.002    207.293      0.000       0.490       0.499
gamma          1.1586      0.002    632.870      0.000       1.155       1.162
==============================================================================
    \end{verbatim}
    \caption{Results for the Poisson Regression of the gravity model (GMI). (statsmodels Python library).}
    \label{tab_poisson}
\end{table}
A comparison between the real flows and the predicted flows of customers from different squares to the malls is shown in the scatter plot in Figure~\ref{scatterp}. The Pearson correlation between real and predicted flows equals $0.67$, and the Spearman correlation equals $0.66$.

\begin{figure}[h!]
\centering
    \includegraphics[height=9cm]{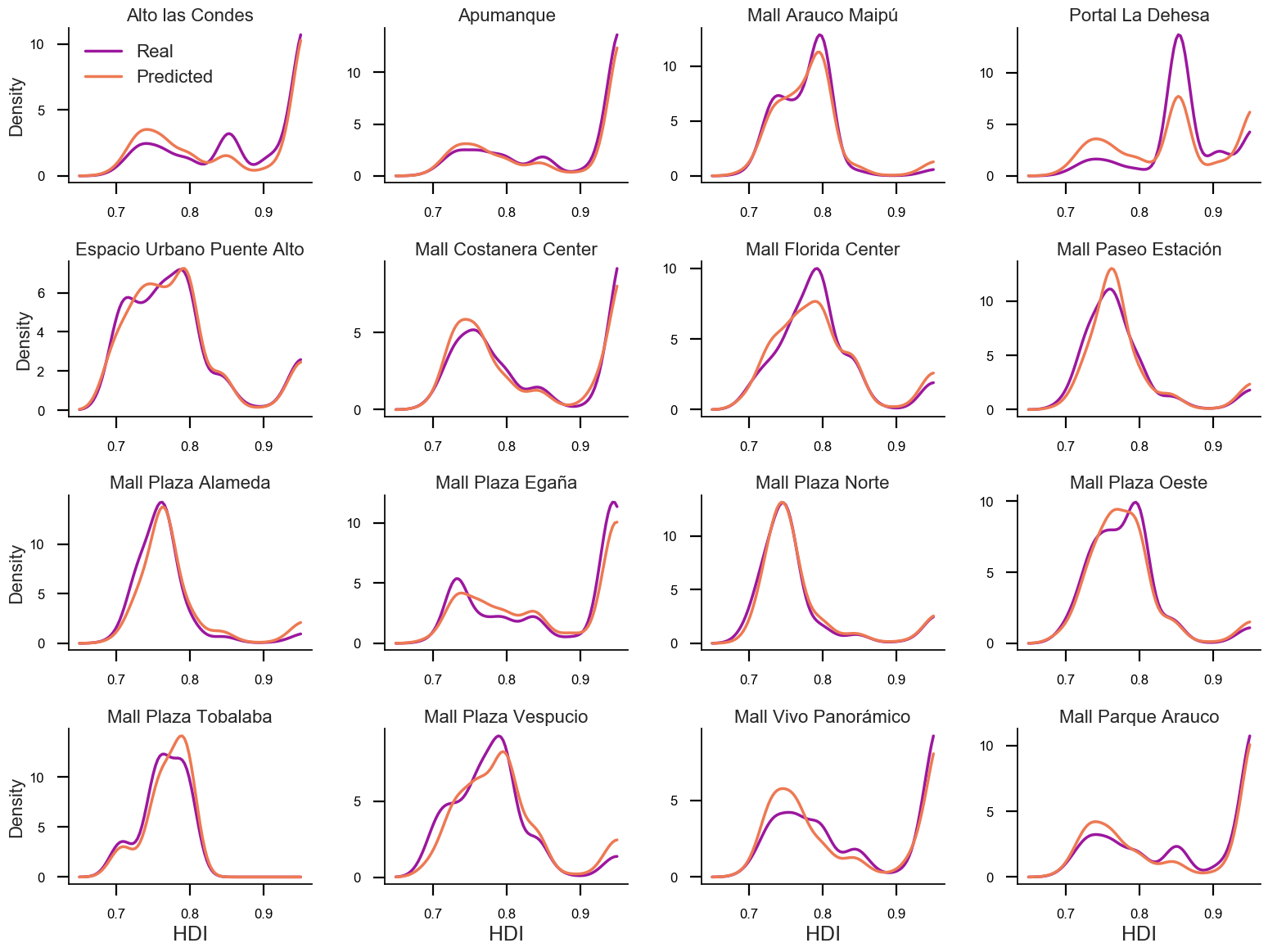}
    \caption{\label{mall_curves}HDI index distribution of customers visiting each mall, according the XDR data (blue) and to the predictions of the fitted gravity model in terms of the influx of people from different \textit{comunas} (red).}
\end{figure}

Now, we are interested in comparing the differences in the profile of customers visiting each mall in the real data and as predicted by the model. We show these differences in Figure~\ref{mall_curves} by plotting a Kernel Density Estimation (KDE) of the HDI indexes of the customers in the real data and in the predicted data. For this purpose, the HDI of the customers arriving from a particular origin square was estimated according to the \textit{comunas} to which each pattern belongs, and the share of each one.

The augmented gravity model 
including the attraction term proved to be overfitting. We do not report the full results, but we obtained a $\lambda$ value of $0.0964 \pm 0.019$ (\textit{i.e.}, people are slightly more attracted by malls with higher customer profiles) with a $z$-value of $5$ (two orders of magnitude below that of the other regressors). This finding implies that social attraction is not significant for mall selection given the other main factors.

\subsection{Co-visitation probability model}

\begin{figure}[h!]
\centering
    \includegraphics[width=12cm]{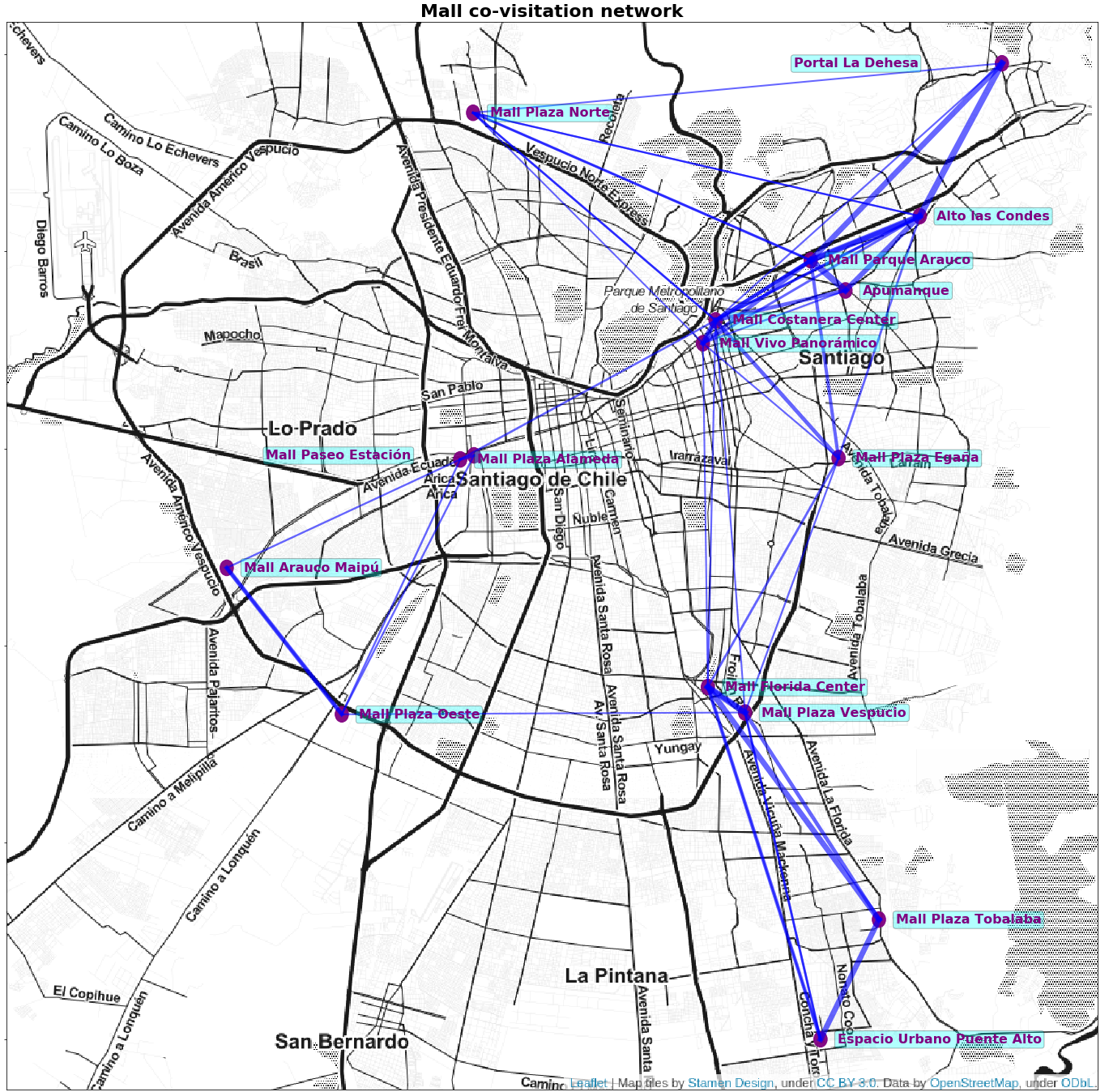}
    \caption{\label{fig:mallpos}Malls network representing the conditional probabilities of visiting one mall during the month, given the fact that another mall is also visited. A minimum threshold of p=0.10 has been applied.}
\end{figure}

\begin{figure}[h!]
\centering
    \includegraphics[height=11cm]{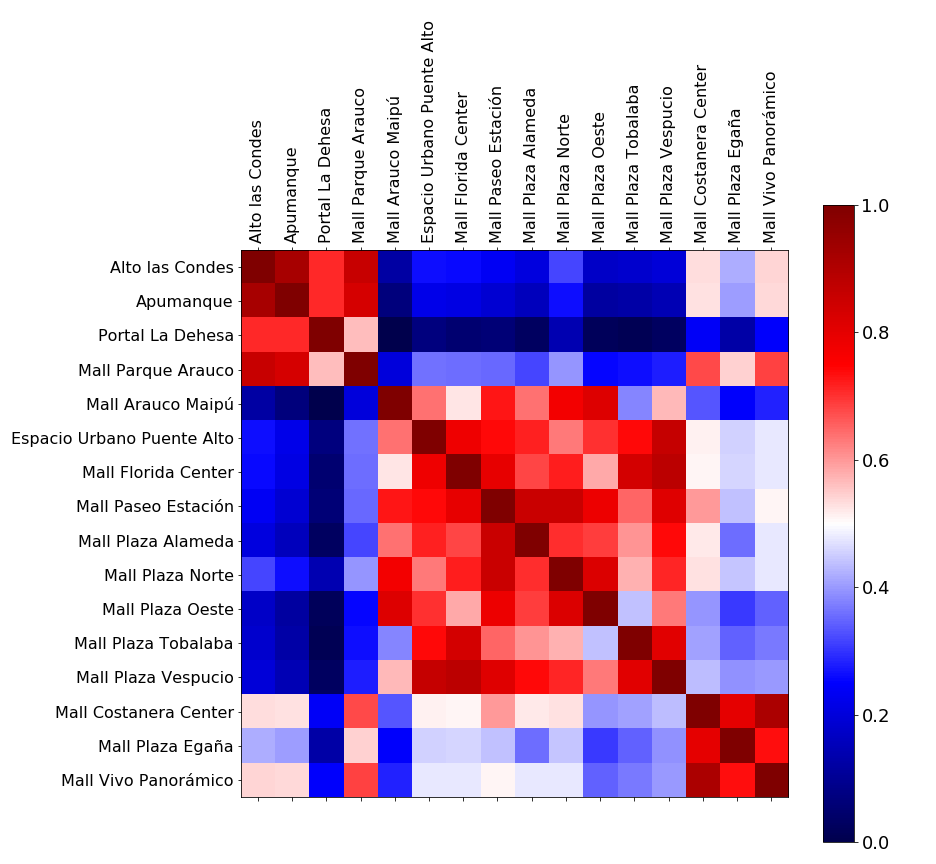}
    \caption{\label{fig_simil_matrix}Similarity matrix between malls. The similarity between a pair of malls was computed as the Kolmogorov-Smirnov distance between between the customer profile distributions of the malls.}
\end{figure}

Figure~\ref{fig:mallpos} shows the network of co-visitations between malls, where the edge-width represents the weight $w$ (the graph is thresholded for $w\geq 0.10$). From here, it seems that co-visitation is also strongly regulated by distance, but we wonder whether people tend to choose similar malls in terms of profile too. In order to run the co-visitation Logit model we computed the similarity between malls as explained in Section~\ref{secMethods}. This similarity matrix is visualized in Figure~\ref{fig_simil_matrix}. The malls in this Figure were sorted according to the $3$ groups that we found by applying a spectral clustering algorithm to the similarity matrix. The cluster (Alto las Condes, Apumanque, Portal La Dehesa, Mall Parque Arauco) is composed by high-target malls, only accessible to people with high HDI. The cluster (Mall Costanera Center, Mall Plaza Egaña, Mall Vivo Panorámico) is composed by three malls in downtown with high mixing level. Finally, the remaining large cluster is composed by 9 malls receiving customers from low and middle classes.

From the results of the co-visitation probability regression model (Table~\ref{tab_poisson2}) we observed that this probability diminishes with distance, as well as it increases with the similarity between malls. Though the coefficients' significance is low, the Spearman correlation between the real and predicted probabilities is $0.75$, and the determination coefficient is $R^2=0.50$.

This result suggests that people tend to visit malls of similar characteristics (though their selection is strongly conditioned by their home location). In fact, a reduced model without considering the similarity term throws a lower determination coefficient of $R^2=0.37$.

\begin{table}[]
    \centering
    \begin{verbatim}
==============================================================================
Dep. Variable:                   p_ij   No. Observations:                   51
Model:                          Logit   Df Residuals:                       47
Method:                           MLE   Df Model:                            3
                                        Pseudo R-squ.:                 0.02800
                                        Log-Likelihood:                -17.508
                                        LL-Null:                       -18.013
                                        LLR p-value:                    0.7992
==============================================================================
                 coef    std err          z      P>|z|      [0.025      0.975]
------------------------------------------------------------------------------
logK          -3.6630      5.644     -0.649      0.516     -14.725       7.399
beta           0.2621      0.500      0.524      0.600      -0.719       1.243
lambda         1.1464      2.035      0.563      0.573      -2.843       5.135
gamma          0.2305      0.522      0.442      0.659      -0.793       1.254
==============================================================================
    \end{verbatim}
    \caption{Results for the Logistic Regression of the co-visitation probability model (CM). (statsmodels Python library).}
    \label{tab_poisson2}
\end{table}

\begin{figure}[h!]
\centering
    \includegraphics[height=7cm]{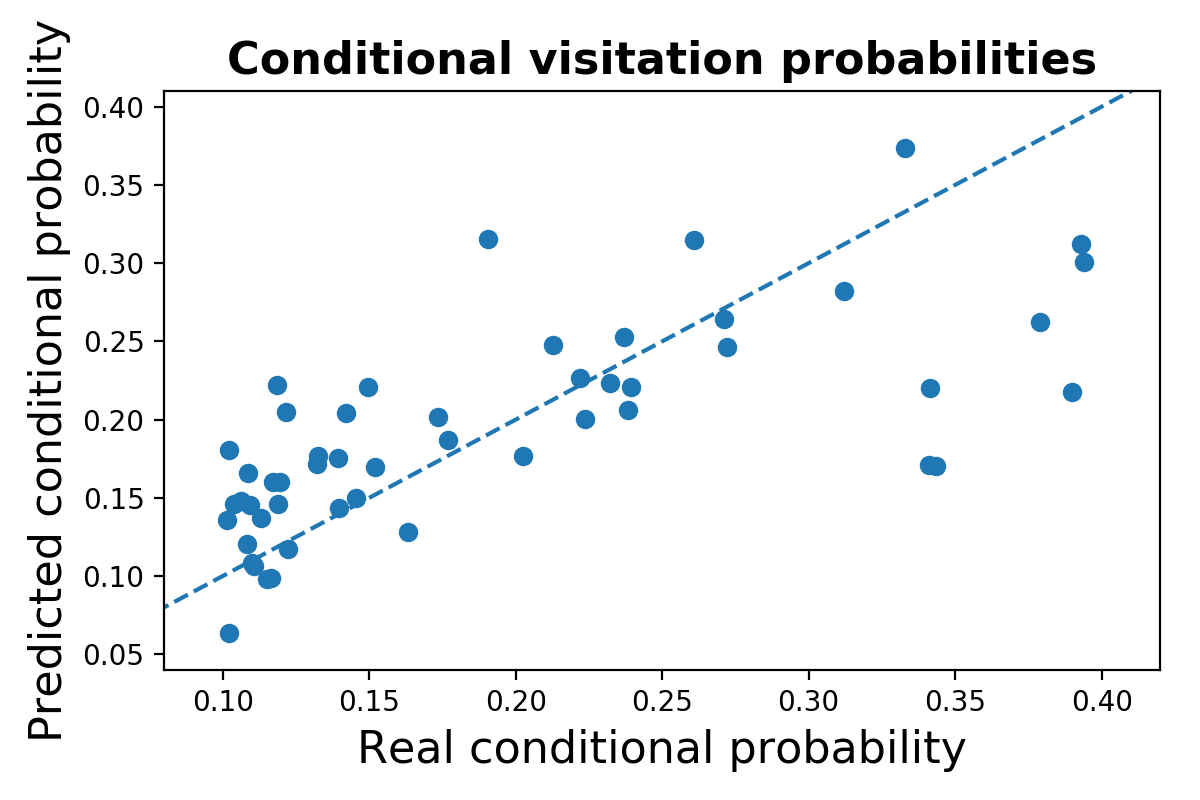}
    \caption{\label{scatterp_prob}Scatter plot representing the predicted conditional co-visitation probabilities against the real ones.}
\end{figure}

\begin{figure}[h!]
\centering
    \includegraphics[height=11cm]{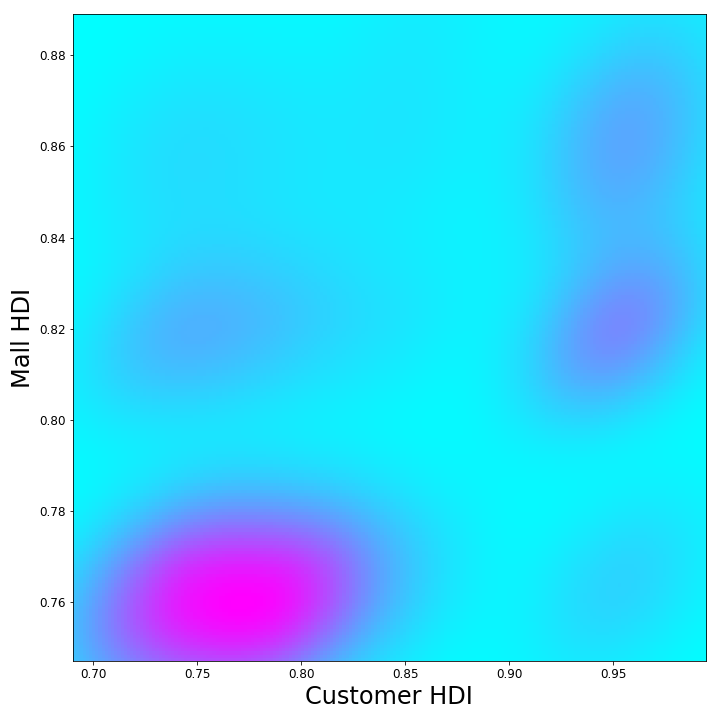}
    \caption{\label{fig_bidimensional}Bidimensional Kernel Density Estimation of the distribution of $\mathit{(customer HDI, mall HDI)}$ where the sample space is the set of all mall visits during the month.}
\end{figure}

Figure \ref{fig_bidimensional} shows a KDE of the distributions of customer HDI against mall HDI (calculated with our method above). One can see a clear effect of visitors of middle-lower classes visiting middle-lower class malls.

\section{Discussion}

According to our findings, the power of malls to \emph{attract} people from diverse social classes may be explained by several factors: every mall, for example, may have a particular target demographics, the time it takes to get to the mall may be such that mall-goers usually choose malls that are nearby their homes (or work), or even that larger malls (in terms of, for instance, leased space) will attract more.

With cellphone data, and given our adjusted gravity model of mall visitation, we have found that the number of people visiting malls, even those coming from different socio-economic backgrounds (given by their home locations) is explained, mostly, by the size and distance to the mall. Meanwhile, the social aspect only explains a tiny fraction of the phenomenon, even though for a mayority of the population malls are a hub of social co-location, which may be a starter for social mixing.

Though the social attraction factor was negligible in our model due to the preponderant effect of distance, this is in part a consequence of the multi-colinearity between distance and social attraction in the regression. In fact, in the co-visitation probability model we found that, given the visit to a specific mall, visits to other malls are influenced by the socio-economic target of the former. This is also expressed in the clustering of malls by customer profile. This lends some evidence to the idea that people have social preferences in their mall choices.

\subsection{Implications}

    Online retail stores (Amazon, EBay, \textit{etc.}) have significantly more information about their customers than their physical counterparts. As online visitors navigate through clicks, compare products, and even give their ``regrets'' (\textit{e.g.}, by emptying the shopping e-carts), they allow for analytics to build a complete profile of the customer base. Physical shopping mall counterparts, on the contrary, do not have such an advantage, and must use the same layout, ads, and deals for everyone. The profiling of mall customers from XDR data done in this work might provide a kick start for improving this situation, making shopping malls more competitive. Note that other data sources could be used in place of XDR data (\textit{i.e.}, Twitter geo-located posts, Swarm check-ins, \textit{etc.}), as our model did not make assumptions about their structure. This makes our model adaptable to different situations and contexts.

\subsection{Scope and Limitations}

As with most C/XDR-based research, there are several limitations that should be acknowledged. Perhaps one of the most important ones have to do with the nature of the dataset itself. Telefónica owns about a third of the Chilean cellphone marketshare. Although the largest, with the second one Entel at about 20\%, this introduces some biases that are hard to identify and quantify. For instance, in its recent history, Telefónica invested heavily on 4G technology, biasing the population of clients towards a higher socio-economic segment. This might be biasing the results obtained and discussed above, in terms of geographic localization, with higher numbers of interactions coming from richer \textit{comunas}.

The calculation of HDI used in this work comprises an integration of several data sources. Most importantly, however, some of these sources belong to different periods of time, the oldest being from 2013 while the most recent being 2015. Thus, the HDI here is a time-range rather than a snapshot of one particular year. We do not consider this to be a major issue here because there has not been a sizeable shift in any of the dimensions measured by the HDI (like education, income or mortality rates) in the \textit{comunas} of Santiago (or all of Chile for that matter) in the period under study. It is, nonetheless, important to mention it here.

A comparison between malls and commercial neighborhoods/streets should be made. That is, there are commercial neighbourhoods in Santiago where social classes might mingle more straightforwardly (such as Patronato\cite{wiki:Patronato} and others). As discussed, malls have a considerable proportion of indoor antennas, which allows us to identify mall visits with a higher degree of certainty, while, conversely, commercial streets may have false positives, \textit{e.g.}, people that pass through but that do not \emph{visit} the place as destination. This is a much more difficult problem that falls out of the scope of this paper.

\subsection{Future Work}

This work can be extended in several dimensions. First of all, it would be interesting to have a more fine-grained description of what kind of mixing happens in those malls where mixing is found (like the Costanera Center). For example, we would like to analyze whether people make a point about going to these malls as day trips during weekends or social mixing may happen only on off-business hours (on leisure time). Another dimension has to do with the role of mall workers, since they may also act as social enablers. A third dimension for future study is people behavior inside malls: since malls have (sometimes more than 100) indoor antennas, we would like to know whether people coming from lower-income \textit{comunas} behave differently inside malls. It will be important to find out whether they visit high-end stores, or simply go to the food-court, for example. Knowing this will help personalize the buying experience of non-target customers, and probably allow \emph{physical} stores compete with online retail stores, with the latter being more fine-tuned to customers and their likes and dislikes.

\section{Conclusion}

At least in Latin America, and in Chile in particular, shopping malls have an important cultural place. Many people visit these places not only for the goods they carry, but to eat, for leisure and in general to interact with other people as well. Usually, less well-off people go to malls because they are free to enter and open to all, are quite safe, and ultimately a nice, comfortable place to spend time in. In this work we looked into the nature of malls as ``social mixers,'' analyzing whether high-end malls can attract low-income visitors. We have done so using XDR data to measure mobility patterns of the people of Santiago, the capital and largest metropolitan area of Chile, where many social classes co-exist.

We found that the mall visitation patterns are predicted mostly by their geographic locations and, relatedly, by distance, and not as much as by socio-economic background, putting to rest some ideas that malls are \textit{inherently} social enablers.

This observed segregation trait notwithstanding, we found that some malls do act as social enablers \textit{de facto} for certain social groups. This is due, in part, to the fact that some malls are in a privileged position in the transportation network, easily accessible from most parts of Santiago. At some point, public policies may be enacted by governments or public organizations to push social mixing, even though it does not happen by itself, and using malls in the same way as public squares were used in the past. All in all, malls are an important landmark of cities. If nothing else, just their sheer volume has a strong impact on the mobility patterns of the people in the cities they're in. We ultimately expect to take advantage of their \emph{attraction} properties to use them as social enablers.


\begin{backmatter}

\section*{Competing interests}
  The authors declare that they have no competing interests.

\section*{Author's contributions}
MB, LF and EG designed the experiments. MB, EG, LF and DC performed data analysis. All author participated in manuscript preparation.

\section*{Acknowledgements}
We thank Telef\'onica R\&D in Santiago for facilitating the data for this study, in particular Pablo García Briosso. We also thank Alonso Astroza for his insightful comments, and Cristi\'an Echeverr\'ia for suggesting the topic of this research.
  
\section*{Availability of data and materials}

The Telef\'onica Movistar mobile phone records have been obtained directly from the mobile phone operator through an agreement between the Data Science Institute and Telef\'onica R\&D. This mobile phone operator retains ownership of these data and imposes standard provisions to their sharing and access which guarantee privacy. Anonymized datasets are available from Telef\'onica R\&D Chile (\url{http://www.tidchile.cl}) for researchers who meet the criteria for access to confidential data. Other datasets used in this study are either derived from mobile records, publicly available, or made available at the dedicated {\tt git} repository: \url{https://github.com/leoferres/mallmob}.


\bibliographystyle{bmc-mathphys} 
\bibliography{bmc_article}      

\end{backmatter}
\end{document}